\documentclass[dvips]{article}
\newcommand{\doublespacing}{\let\CS=\@currsize\renewcommand{\baselinesstrech}
{2.0}\tiny\CS}

\begin{document}

\textwidth 16cm
\newcommand{\bd}{\begin{document}}
\newcommand{\ed}{\end{document}}
\newcommand{\bc}{\begin{center}}
\newcommand{\ec}{\end{center}}
\newcommand{\bfr}{\begin{flushright}}
\newcommand{\efr}{\end{flushright}}
\newcommand{\lt}{\left}
\newcommand{\rt}{\right}
\newcommand{\vs}{\vspace}
\newcommand{\hs}{\hspace}
\newcommand{\beq}{\begin{equation}}
\newcommand{\eeq}{\end{equation}}
\newcommand{\lb}{\linebreak}
\newcommand{\pb}{\pagebreak}
\newcommand{\mb}{\makebox}
\newcommand{\fb}{\framebox}
\newcommand{\mc}{\multicolumn}
\newcommand{\ben}{\begin{enumerate}}
\newcommand{\een}{\end{enumerate}}
\newcommand{\bit}{\begin{itemize}}
\newcommand{\eit}{\end{itemize}}
\newcommand{\ol}{\overline}
\newcommand{\un}{\underline}
\newcommand{\lefq}{\lefteqn}
\newcommand{\ba}{\begin{array}}
\newcommand{\ea}{\end{array}}
\newcommand{\beqa}{\begin{eqnarray}}
\newcommand{\eeqa}{\end{eqnarray}}
\newcommand{\beqas}{\begin{eqnarray*}}
\newcommand{\eeqas}{\end{eqnarray*}}
\newcommand{\bfg}{\begin{figure}}
\newcommand{\efg}{\end{figure}}
\newcommand{\bds}{\begin{displaymath}}
\newcommand{\eds}{\end{displaymath}}
\newcommand{\btb}{\begin{tabbing}}
\newcommand{\etb}{\end{tabbing}}
\newcommand{\para}{\parallel}
\newcommand{\pad}{\partial}
\newcommand{\nn}{\nonumber}
\newcommand{\la}{\leftarrow}
\newcommand{\ra}{\rightarrow}
\newcommand{\lgla}{\longleftarrow}
\newcommand{\lgra}{\longrightarrow}
\newcommand{\La}{\Leftarrow}\newcommand{\Ra}{\Rightarrow}
\newcommand{\Lra}{\Leftrightarrow}
\newcommand{\Lgla}{\Longleftarrow}
\newcommand{\Lgra}{\Longrightarrow}
\newcommand{\bm}{\boldmath}
\newcommand{\lan}{\langle}
\newcommand{\ran}{\rangle}
\renewcommand{\a}{\alpha}
\renewcommand{\b}{\beta}
\newcommand{\g}{\gamma}
\newcommand{\G}{\Gamma}
\renewcommand{\d}{\delta}
\newcommand{\eps}{\epsilon}
\newcommand{\Th}{\Theta}
\newcommand{\s}{\sigma}
\newcommand{\lam}{\lambda}
\newcommand{\D}{\Delta}
\newcommand{\vare}{\varepsilon}
\newcommand{\pr}{\prime}
\newcommand{\ro}{\rho}
\newcommand{\nab}{\nabla}
\newcommand{\m}{\mu}
\newcommand{\n}{\nu}
\newcommand{\Sg}{\Sigma}
\newcommand{\p}{\pi}
\newcommand{\R}{I\!\!R}
\newcommand{\om}{\omega}
\newcommand{\Om}{\Omega}
\newcommand{\ze}{\zeta}
\newcommand{\vart}{\vartheta}
\newcommand{\tri}{\triangle}
\newcommand{\f}{\frac}
\newcommand{\iny}{\infty}
\newcommand{\pro}{\propto}
\bc {\huge Construction of ${\cal{C}}$ operator for a ${\cal{PT}}$ symmetric model} \ec

\vs{1cm}

\bc
{\it R. Roychoudhury{\footnote {e-mail : raj@isical.ac.in} and P. Roy{\footnote{e-mail : pinaki@isical.ac.in}}\\
Physics \& applied Mathematics Unit \\
Indian Statistical Institute \\
Kolkata - 700 108, India.}} \ec
\vs{4.5cm}

\bc {\large {\un{abstract}}} \ec 
We obtain a closed form expression of the ${\cal{C}}(x,y)$ operator for the $\cal{PT}$ symmetric Scarf I potential. It is also shown that the eigenfunctions form a complete set.\\
\pb

In recent years non Hermitian systems, in particular the $\cal{PT}$ symmetric ones \cite{bender} have been studied widely. Many of these systems are characterised by the fact that they possess real eigenvalues. However for non Hermitian systems the concept of a scalar product is a non trivial one. In fact a straight forward $\cal{PT}$ symmetric generalisation of the usual scalar product for Hermitian systems produces a norm which alternates in sign i.e,
\beq
<\psi_m|\psi_n>_{\cal{PT}} = (-1)^n\delta_{mn}\label{ortho}
\eeq
With a view to circumvent this difficulty an operator ${\cal{C}}(x,y)$ was introduced \cite{benderc}. This operator is defined as \cite{benderc}
\beq
{\cal{C}}(x,y) = \sum_{n=0}^\infty \psi_n(x)\psi_n(y)\label{c}
\eeq
where $\psi_n(x)$ are eigenfunctions of the Hamiltonian $H$:
\beq
H\psi_n(x) = \lambda_n \psi_n(x)
\eeq
However, it is not always easy to obtain a closed form expression of the ${\cal{C}}(x,y)$ operator and often one has to construct it using various approximating techniques \cite{benderc1}. Our purpose here is to obtain a closed form expression of the ${\cal{C}}(x,y)$ operator for a $\cal{PT}$ symmetric Scarf I potential. 
\vspace{.2cm}

We consider the Scarf I potential defined by
\beq
V(x) = \left(\f{\alpha^2+\beta^2}{2}-\f{1}{4}\right)\f{1}{cos^2x}+\f{\alpha^2-\beta^2}{2}\f{sinx}{cos^2x}~~,~~x\in [-\f{\pi}{2},\f{\pi}{2}]\label{scarf1}
\eeq
where $\alpha$ and $\beta$ are complex parameters such that $\beta^*=\alpha$ and $\alpha_R>\f{1}{2}$. In this case the (real) eigenvalues and the corresponding eigenfuctions are given by \cite{levai}
\beq
\ba{lcl}
E_n &=& \left(n + \f{\alpha+\beta+1}{2}\right)^2\\
\psi_n(x) &=& D_n (1-sin x)^{\f{\alpha}{2}+\f{1}{4}} (1+sin x)^{\f{\alpha^*}{2}+\f{1}{4}} P_n^{(\alpha,\alpha^*)}(sinx),~~~~n=0,1,2,......\label{wf}
\ea
\eeq
where $P_n^{(a,b)}(x)$ denotes the Jacobi polynomial and $D_n$ is a normalisation constant given by
\beq
D_n = i^n \sqrt{\f{(2n+2\alpha_R+1)n!\Gamma(n+2\alpha_R+1)}{2^{2\alpha_R+1}\Gamma(n+\alpha+1)\Gamma(n+\alpha^*+1)}}
\eeq
Using the orthogonality properties of Jacobi polynomials \cite{abram} it can be shown \cite{levai} that the wave functions in (\ref{wf}) satisfy the relation
\beq
\int_{-\pi/2}^{\pi/2}({\cal{PT}}\psi_m(x))\psi_n(x)~dx = (-1)^n\delta_{mn}\label{ortho1}
\eeq
We now turn to the evaluation of the ${\cal{C}}(x,y)$ operator. Using (\ref{wf}) we obtain from (\ref{c})
\beq
{\cal{C}}(x,y) = \prod_{z=x,y}(1-sin z)^{\f{\alpha}{2}+\f{1}{4}} (1+sin z)^{\f{\alpha^*}{2}+\f{1}{4}} \sum_{n=0}^\infty \f{(-1)^n(2n+2\alpha_R+1)n!\Gamma(n+2\alpha_R+1)}{{2^{2\alpha_R+1}\Gamma(n+\alpha+1)\Gamma(n+\alpha^*+1)}} P_n^{(\alpha,\alpha^*)}(sinx)P_n^{(\alpha,\alpha^*)}(siny)\label{c1}
\eeq
To evaluate the summation in (\ref{c1}) we now use the result \cite{hari}
\beq
\sum_{n=0}^\infty n!\f{(2\alpha_R+1)_n}{(\alpha+1)_n(\beta+1)_n}(2n+2\alpha_R+1)P_n^{(\alpha,\alpha^*)}(sinx)P_n^{(\alpha,\alpha^*)}(siny)t^n = \f{(2\alpha_R+1)(1-t)}{(1+t)^{2\alpha_R+1}}~F_4(a,b,c,d,U,V)\label{sum1}
\eeq
where 
\beq
F_4(a,b,c,d,U,V) = \sum_{r,s=0}^\infty \f{(a)_s(b)_s}{s!(d)_s}\f{(a+s)_r(b+s)_r}{r!(c)_r}~U^rV^s = \sum_{s=0}^\infty \f{(a)_s(b)_sV^s}{s!(d)_s }~{_2F_1}(a+s,b+s,c,U)\label{sum2}
\eeq
\beq
a = \alpha_R+1,~~b = \alpha_R+3/2,~~c = 1+ \alpha,~~d = \beta +1,~~U = \f{(1-sinx)(1-siny)t}{(1+t)^2},~~V = \f{(1+sinx)(1+siny)t}{(1+t)^2}
\eeq
and ${_2F_1(a,b,c,z)}$ is the standard hypergeometric function.

Now taking the limit $t\rightarrow -1$, we obtain
\beq
F_4(a,b,c,d,U,V) = (-U)^a\sum_{s=0}^\infty \f{\Gamma(c)\Gamma(1/2)}{\Gamma(b+s)\Gamma(c-a-s)}\f{(-V/U)^s(a)_s(b)_s}{s!(d)_s}\label{f4}
\eeq
Then using (\ref{f4}) we obtain from (\ref{sum1}) and (\ref{sum2})
\beq
{\cal{C}}(x,y) = {\cal{N}}\f{[(1+sinx)(1+siny)]^{(\alpha^*/2+1/4)}}{[(1-sinx)(1-siny)]^{(\alpha^*/2+3/4)}}~{_2F_1(a,1-c+b,d,z)},~~~~z = \f{(1+sinx)(1+siny)}{(1-sinx)(1-siny)}\label{c3}
\eeq
where $\cal{N}$ is a constant given by
\beq
{\cal{N}} = \f{2\Gamma(\alpha_R+1)sin(\pi(1-c+a))\Gamma(1-c+a)}{\pi~\Gamma(\alpha^*+1)}
\eeq
It may be noted that (\ref{c3}) is an exact result. 
\vspace{.25cm}

{\bf{Completeness of the eigenfunctions}}

The completeness property is very important feature of eigenfunctions. However, to the best of our knowledge for $\cal{PT}$ symmetric systems this property has been verified numerically \cite{bender3}. Here we shall show analytically that the eigenfunctions (\ref{wf}) form a complete set. To do this we note that in a $\cal{PT}$ symmetric theory with unbroken $\cal{PT}$ symmetry the completeness property can be expressed as \cite{benderc,benderc1}
\beq
\sum_{n=0}^\infty (-1)^n \psi_n(x)\psi_n(y) = \delta (x-y)\label{complete}
\eeq
To prove (\ref{complete}) we consider the result \cite{wolfram}
\beq
\sum_{n=0}^\infty \f{n!\Gamma(a+b+2n+1)\Gamma(a+b+n+1)}{\Gamma(a+n+1)\Gamma(b+n+1)}~P_n^{(a,b)}(x)P_n^{(a,b)}(y) = (1+x)^{-b/2}(1-x)^{-a/2}(1+y)^{-b/2}(1-y)^{-a/2}\delta(x-y)\label{com}
\eeq
where $-1<x,y<1, Re(a)>-1, Re(b)>-1$. Now putting $a=\alpha, b=\beta$ in (\ref{com}) and using (\ref{wf}) we obtain
\beq
\sum_{n=0}^\infty (-1)^n \psi_n(x)\psi_n(y) = {\sqrt{cosx~cosy}}~\delta (sinx-siny) = \delta(x-y)
\eeq
Thus the eigenfunctions (\ref{wf}) form a complete set. 

It is interesting to note that two important results can be derived using (\ref{com}). First we recall that in Hermitian systems, the operator ${\cal{C}}(x,y)$ is just the parity operator i.e., ${\cal{C}}(x,y) = \delta (x+y)$. So for $\alpha = \alpha^*$, (\ref{c3}) should reduce to this limit. Now using the properties of Hypergeometric functions it can be shown that for real $\alpha,\beta$$, {\cal{C}}(x,y) = \delta(x+y)$. The other properties of the ${\cal{C}}$ operator viz, ${\cal{C}}\psi_n = (-1)^n\psi_n$ follow from the definition (\ref{c}) and (\ref{ortho}) while ${\cal{C}}^2 = 1$ can be derived using the results (\ref{ortho1}) and (\ref{com}).
 
\vspace{.25cm}

{\center
{\bf{Acknowledgement}}}

One of the authors (RR) is grateful to the Council of Scientific and Industrial Research  (CSIR) for financial support (Project no 21/(0659)/06/EMR-II ).

\ed